# The pH of Enceladus' ocean

Christopher R. Glein [a*], John A. Baross [b], and J. Hunter Waite Jr. [c]

[a] Geophysical Laboratory, Carnegie Institution of Washington,
5251 Broad Branch Road NW, Washington, DC 20015-1305, United States
Phone: 202-478-8967, Fax: 202-478-8901, Email: cglein@ciw.edu

[*] Present address: Department of Earth Sciences, University of Toronto,
Earth Sciences Centre, 22 Russell Street, Toronto, ON M5S 3B1, Canada
Phone: 416-978-5508, Fax: 416-978-3938, Email: chris.glein@utoronto.ca

[b] Astrobiology Program and School of Oceanography, University of Washington
[c] Space Science and Engineering Division, Southwest Research Institute

# ABSTRACT

Saturn's icy moon, Enceladus, is a geologically active waterworld. The prevailing paradigm is that there is a subsurface ocean that erupts to the surface, which leads to the formation of a plume of vapor and ice above the south pole. The chemical composition of the ocean is just beginning to be understood, but is of profound geochemical and astrobiological interest. Here, we address the most fundamental question about the ocean's chemistry by determining its pH, using a thermodynamic model of carbonate speciation. Observational data from the *Cassini* spacecraft are used to obtain a chemical model of ocean water on Enceladus, the first such model for another planet/moon that is based on empirical evidence. The abundance of $CO_2$ in the plume gas, as measured by the Ion and Neutral Mass Spectrometer onboard *Cassini*, is essential to elucidating the pH. The geochemical model indicates that Enceladus' ocean is a Na-Cl-$CO_3$ solution with an alkaline pH of ~11-12. The dominance of aqueous NaCl is a geochemical feature that Enceladus' ocean shares with terrestrial seawater, but the ubiquity of dissolved $Na_2CO_3$ suggests that soda lakes are more analogous to the Enceladus ocean. The high pH implies that the hydroxide ion should be relatively abundant, while divalent metals should be present at low concentrations in ocean water owing to buffering by clays and carbonates on the ocean floor. In addition, carboxyl groups in dissolved organic species (such as possible biomolecules) would be negatively charged, whereas amino groups would exist predominately in the neutral form. Knowledge of the pH dramatically improves our understanding of geochemical processes in Enceladus' ocean. In particular, the high pH is interpreted to be a key consequence of serpentinization of chondritic rock, as predicted by prior geochemical reaction path models, although degassing of $CO_2$ from the ocean may also play a role depending on the efficiency of mixing processes in the ocean. Serpentinization inevitably leads to the generation of $H_2$, a geochemical fuel that can support both abiotic and biological synthesis of organic molecules such as those that have been detected in Enceladus' plume. Serpentinization and $H_2$ generation should have occurred on Enceladus, like on the parent bodies of aqueously altered meteorites; but it is unknown whether these critical processes are still taking place, or if Enceladus' rocky core has been completely altered by past hydrothermal activity. The presence of native $H_2$ in the plume would provide strong evidence for contemporary aqueous alteration. The high pH also suggests that the delivery of oxidants from the surface to the ocean has been sporadic, and the rocky core did not experience partial melting and igneous differentiation. On the other hand, the deduced pH is completely compatible with life as we know it; indeed, life on Earth may have begun under similar conditions, and terrestrial serpentinites support thriving microbial communities that are centered on $H_2$ that is provided by water-rock reactions. These considerations provide major motivation for future missions to explore Enceladus as a habitable world, whether past or present.

# 1. INTRODUCTION

A wealth of geological (Smith et al., 1982; Porco et al., 2006; 2014; Spencer et al., 2009), geophysical (Spencer et al., 2006; Schmidt et al., 2008; Howett et al., 2011; Ingersoll & Ewald, 2011; Iess et al., 2014), and geochemical (Postberg et al., 2009; 2011; Waite et al., 2009; Hsu et al., 2014) evidence points to the presence of a global or regional ocean of liquid water beneath the icy crust of the south polar region of Saturn's satellite, Enceladus (Collins & Goodman, 2007; Nimmo et al., 2007; Tobie et al., 2008; Běhounková et al., 2012). This ocean is believed to be the source of gases, organics, salts, and ices in Enceladus' cryovolcanic plume (Spencer & Nimmo, 2013), although clathrate hydrates could also play a role (Kieffer et al., 2006; Fortes, 2007; Bouquet et al., 2014). Understanding the geochemistry of the ocean is critical for advancing our understanding of the origin and evolution of Enceladus (Zolotov, 2007; Matson et al., 2007; Glein et al., 2008; Zolotov et al., 2011); for gaining insights into geochemical processes on other icy bodies in the outer Solar System (Shock & McKinnon, 1993; Kargel et al., 2000; Glein et al., 2009); for comparing aqueous geochemistry on Enceladus to that on the modern and early Earth (Holland, 1984; Sverjensky & Lee, 2010; Pope et al., 2012), Mars (Grotzinger et al., 2014), and asteroids (Zolensky et al., 1999; Brearley, 2006; Zolotov, 2014); for assessing the biological potential of Enceladus (McKay et al., 2008; 2014; Parkinson et al., 2008); and for planning future missions to Enceladus (Sotin et al., 2011; Tsou et al., 2012) and other worlds where ocean water may be erupting into space (Pappalardo et al., 2013; 2014; Küppers et al., 2014; Roth et al., 2014).

The most important compositional variable to understanding the fundamental low-temperature geochemistry of natural waters is the pH (Garrels & Christ, 1965; Drever, 1997), which reflects the acidity of the water and is formally a measure of the thermodynamic activity of the hydrogen ion (pH = −log[$a_{H+}$]) referenced to the infinitely dilute standard state [1] (the second most important variable is the oxidation state; e.g., the reduction potential, Eh). The pH of Enceladus' ocean is currently unknown, although predictions have been made based on the assumption of chemical equilibrium between chondritic rock and water (Zolotov, 2007; 2012). However, the robustness of these models to icy moons has not been tested. Postberg et al. (2009) deduced the existence of $NaHCO_3$ and/or $Na_2CO_3$ in plume particles, and pointed out that their presence implies a basic pH; but they were unable to pinpoint the pH because of the uncertain carbonate speciation (i.e., $HCO_3^-$ vs. $CO_3^{-2}$). They did suggest a pH of 8.5-9 based on patterns in their mass spectra, but they did not explain how they came to this conclusion, nor did they provide a discussion of the uncertainty associated with this value. Recently, it has been argued that the formation of silica nanoparticles inside Enceladus requires a pH of 8-10 (Hsu et al., 2014). However, this interpretation is based on a hydrothermal scenario for the formation of the nanoparticles (Sekine et al., 2014) that may be geochemically implausible (Zolotov & Postberg, 2014). Direct measurement of the pH of Enceladus' ocean may be possible in the future, but will require a complex and costly drilling mission to access the liquid.

Fortunately, a synthesis of chemical data from the *Cassini* spacecraft allows an indirect yet rigorous determination of the pH of Enceladus' ocean. Indeed, the primary objective of the present paper is to show how *in situ* observations can be used to estimate the pH of an extraterrestrial ocean by linking the latest measurements of carbonate salts in the plume particles (Postberg et al., 2009; 2011) and $CO_2$ in the plume gas (Waite et al., 2011; 2013) with a thermodynamic model of the classic carbonate system at the conditions of Enceladus' ocean (it should be noted that Marion et al. [2012] attempted to estimate the pH of the ocean in a similar manner, but they adopted an unrealistically high partial pressure of $CO_2$ as an initial constraint [see Section 2.2]). Knowledge of the pH allows a number of important inferences to be made regarding geochemical processes in Enceladus' interior, given that terrestrial experience (Garrels & Christ, 1965; Drever, 1997) demonstrates that the pH of natural waters is controlled by and thus reflects geochemical processes. The pH is also very important to the habitability of Enceladus' ocean because the composition of the environment sets the boundary conditions for metabolism and other life processes.

---

[1] Conventional thermodynamic standard states (Anderson, 2005) are adopted in the present communication. These correspond to: unit activity of pure liquid water at any temperature and pressure, unit activity of other aqueous species in a hypothetical one molal solution referenced to infinite dilution at any temperature and pressure, and unit fugacity of the hypothetical ideal gas at 1 bar and any temperature.

This paper is structured as follows. In the following section, we derive constraints on the geochemical conditions of Enceladus' ocean from *Cassini* data and published interpretations of the data. We then present illustrative (Section 3.1) and realistic (Section 3.2) equilibrium models for the pH and speciation of the ocean. Subsequently, we discuss what the results mean in terms of geochemical processes on Enceladus (Section 4). In the final section, we consider the astrobiological implications, particularly the prospects for life.

## 2. GEOCHEMICAL CONSTRAINTS

### 2.1. System and conditions

Enceladus presents a unique variation on the well-known problem of carbonate equilibria (Garrels & Christ, 1965). Here, it is assumed that an ocean is the source of the ice particles and gases in Enceladus' plume, as supported by prior studies (Porco et al., 2006; 2014; Schmidt et al., 2008; Postberg et al., 2009; 2011; Waite et al., 2009; Ingersoll & Ewald, 2011). Mass spectra obtained by the Cosmic Dust Analyzer (CDA) onboard *Cassini* show that NaCl and $NaHCO_3/Na_2CO_3$ salts are the most abundant non-water constituents in the plume particles (Postberg et al., 2009; 2011). $K^+$ has also been detected, but is much less abundant than $Na^+$ (Postberg et al., 2009). Silicon in Saturnian stream particles (Kempf et al., 2005) has been inferred to be derived from Enceladus' plume (Hsu et al., 2014), but the concentration of $SiO_2$ in the plume particles is presently unknown. The plume gas is dominated by water vapor (Hansen et al., 2006; 2011), but data from the Ion and Neutral Mass Spectrometer (INMS) show that it also contains appreciable quantities of $CO_2$ (Waite et al., 2006; 2009; 2011; 2013), which can explain the presence of solid $CO_2$ in the south polar region (Brown et al., 2006). The other reported plume gases (e.g., $CH_4$, $NH_3$) are not considered here because they do not provide significant constraints on the pH of Enceladus' ocean ($NH_3$ could if the $NH_4^+$ concentration can be determined; Postberg et al., 2009).

The simplest geochemical system that is consistent with the preceding observations is $Na_2O$-$HCl$-$CO_2$-$H_2O$. According to Gibbs' phase rule, the number of degrees of freedom that are required to define thermodynamic equilibrium is: $F = C-P+2$, where $C$ stands for the number of components (= 4), and $P$ represents the number of phases present (i.e., the ocean). Therefore, $F = 5$ for this system. The most convenient intensive variables for this particular problem are temperature, total pressure, the activity of $CO_2$, and the concentrations of chloride and dissolved inorganic carbon in the ocean. The remainder of this section focuses on constraining these quantities from *Cassini* data.

Setting the temperature and total pressure is straightforward. The ocean will be covered by a layer of water ice (Brown et al., 2006; Iess et al., 2014), so it can be assumed that the two phases are in thermal equilibrium. Therefore, the temperature should be close to 0°C unless the ocean is extremely salty, which seems unlikely based on measurements of plume particles (Postberg et al., 2009; 2011). The plume also does not contain sufficient $NH_3$ (Waite et al., 2009; 2011; 2013) to depress the freezing point significantly (Marion et al., 2012). The plume gas is envisioned to form when tidally-opened (Hedman et al., 2013; Nimmo et al., 2014) cracks expose the ocean to low pressure (Schmidt et al., 2008). The newly formed water vapor will be in equilibrium with the ocean and will provide the dominant source of pressure owing to the low abundances of other gases in Enceladus' plume (Waite et al., 2009; 2011; 2013). As a result, the total pressure at the interface between the ocean and plume gas should be similar to the vapor pressure of water at 0°C (6.11 mbar; Haynes, 2014). Here, thermodynamic calculations are performed at a total pressure of 1 bar, because this is the reference pressure at which equilibrium constants are commonly tabulated (Bethke, 2008), and pressure effects on the equilibrium constants are negligible at these low pressures (Marion et al., 2005).

## 2.2. Carbon dioxide

The activity of $CO_2$ can be calculated if it is assumed that the $CO_2$ in the plume gas was in equilibrium with ocean water in the subsurface (Matson et al., 2012). According to Henry's law: $K_H = a_{CO2}/f_{CO2}$; where $K_H$ corresponds to the equilibrium constant (= $10^{-1.1}$ at 0°C and 1 bar; Shock et al., 1989), $a_{CO2}$ denotes the activity of aqueous $CO_2$ (often referred to as carbonic acid in the geochemical literature), and $f_{CO2}$ stands for the fugacity of $CO_2$. At the low total pressure of interest, the fugacity will be essentially identical to the partial pressure (Anderson, 2005). Thus, the present problem is to constrain the partial pressure of $CO_2$ at the ocean-vapor interface inside Enceladus. The situation is analogous to carbonate equilibria in surface waters on Earth (e.g., rainwater), where the partial pressure of atmospheric $CO_2$ controls its aqueous activity (Drever, 1997). In this context, the plume of Enceladus can be thought of as its "atmosphere". However, the situation is actually more complicated because the plume is expanding into the vacuum of space (Hansen et al., 2006), meaning that the partial pressure in the plume will be dramatically lower than at the ocean-vapor interface.

A solution to this problem is to calculate the partial pressure of $CO_2$ at depth using the molar ratio of $CO_2/H_2O$ via Dalton's law: $p_{CO2} = (CO_2/H_2O) \times p_{H2O}$, where $p_i$ refers to the partial pressure of species $i$. The partial pressure of steam at the surface of the ocean can be taken as 6.11 mbar (see Section 2.1). A new complication, however, is that the $CO_2/H_2O$ ratio at the ocean-vapor interface cannot be assumed to be the same as that in the plume gas, because steam will condense as its travels through cracks connecting the warm ocean and cold surface of Enceladus (Schmidt et al., 2008; Porco et al., 2014). The implication is that the $CO_2/H_2O$ ratio will be higher in the plume gas than in the source region. This is an important point because it has been assumed that the composition of the plume can simply be equated to that of the source region (Kieffer et al., 2006; Waite et al., 2009). Assuming phase equilibrium between steam and ice (Ingersoll & Pankine, 2010), the following mass balance can be used to estimate the $CO_2/H_2O$ ratio in the source region by accounting for condensation of water vapor during the transport process from the bottom (ocean) to the top (tiger stripes; Spitale & Porco, 2007) of the cracks

$$\rho_{ocean}^{steam} \times (CO_2/H_2O)_{ocean} = \rho_{tiger}^{steam} \times (CO_2/H_2O)_{tiger}, \qquad (1)$$

where $\rho^{steam}$ designates the saturation molar density of water vapor at the ocean or tiger stripes, and the $CO_2/H_2O$ ratio at the tiger stripes is taken to be equal/similar to that in the plume gas, as expansion of the gas into a vacuum at the surface should not alter the ratio because no work is performed, and the formation of appreciable amounts of ice in the plume seems unlikely because there would not be enough collisions to nucleate ice in the dilute plume (Schmidt et al., 2008; Hansen et al., 2011), although some ice in the form of nanograins (Jones et al., 2009) may condense at the base of the plume where sufficient collisions occur (Yeoh et al., 2013). This would be important to Eq. (1) if a significant fraction of the ice in the plume exists as nanograins that formed at/above the surface, and if the nanograins are chemically fractionated from the plume gas (i.e., they have different $CO_2/H_2O$ ratios). This possibility is poorly understood at present, so we will neglect it for the time being; however, dynamical modeling (Yeoh et al., 2013) may allow a quantitative assessment of its importance in the future. Equation (1) relies upon the assumption that water vapor condenses in the cracks (Schmidt et al., 2008) while $CO_2$ gas does not, which is probably not completely correct given the reported detection of $CO_2$ ice in the tiger stripes (Brown et al., 2006). Nevertheless, this equation represents a suitable endmember scenario that reflects the much greater propensity of steam to condense than $CO_2$, so it should provide a reasonable approximation to the more complex process on Enceladus (see Section 3.2.2.4).

For an ocean temperature of 0°C (273.15 K), it is found that the ratio of steam densities can be parameterized as: $\log(\rho_{tiger}/\rho_{ocean}) = 9.429 - 2574/T_{tiger}(K)$, which represents the experimentally determined density ratio (Haynes, 2014) to within ~2% from 153-273 K.

Combining the previous equations then leads to an expression for the activity of $CO_2$ in the ocean as a function of the $CO_2/H_2O$ ratio in the plume, and the temperature of the tiger stripes

$$\log(a_{CO_2}) = \log(CO_2/H_2O)_{plume} + 6.115 - 2574/T_{tiger}(K). \qquad (2)$$

Recently, INMS has been flown through the plume several times at lower velocities, which has led to a revision of the composition of the plume gas (Waite et al., 2011; 2013). The slower flybys give repeatable results, and provide a better indication of the "true" plume composition because collisionally induced chemical modification of the plume gas becomes less important. Most critical to this work is that the $CO_2/H_2O$ ratio has been revised significantly from 6% (Waite et al., 2009) to 0.6% (Waite et al., 2013). The latter value is regarded as the most accurate one, and is adopted here.

While we presently lack fine-scale information on the temperature distribution of all four tiger stripes, it seems reasonable to assume that average plume material is erupted at a temperature similar to that of Baghdad Sulcus (200±20 K; Goguen et al., 2013), since this tiger stripe appears to be the biggest contributor to the plume (Spitale & Porco, 2007). In light of the uncertainty, however, it would seem prudent to adopt a more conservative temperature range of 170-230 K. Inserting this and $CO_2/H_2O = 0.006$ into Eq. (2) leads to the constraint that $\log(a_{CO2})$ should be between $-11.2$ and $-7.3$. The corresponding partial pressure of $CO_2$ is $10^{-10.1}$-$10^{-6.2}$ bar, which is much lower than the value (0.349 bar) adopted by Marion et al. (2012), who did not account for enrichment of $CO_2$ in the plume by deposition of water vapor during plume formation (i.e., Eq. 1), and they assumed that the ocean is in equilibrium with a $CO_2$ clathrate hydrate. The derived low activity (~molality) of $CO_2$ is also inconsistent with the "Perrier

Ocean" model of Matson et al. (2012), which requires much more $CO_2$ to drive bubble formation.

## 2.3. Chloride and carbonate

Geochemical and geophysical approaches can be used to constrain the chloride concentration in Enceladus' ocean. Postberg et al. (2009; 2011) identified peaks in CDA mass spectra of salt-rich plume particles that they attributed to NaCl-bearing clusters. They also attempted to reproduce the CDA data by performing laboratory experiments with a laser, and found that solutions with 0.05-0.2 mol of NaCl per kg of water can produce spectra that are remarkably consistent with those from Enceladus (Postberg et al., 2009). Therefore, this represents a possible range for the chloride concentration in the ocean, if the salt-rich plume particles are formed from flash freezing of aerosolized ocean water, which seems like the most likely explanation for the existence of salts in small ice particles (Spencer & Nimmo, 2013). However, there are potential caveats that may preclude direct application of these laboratory data to Enceladus. First, water vapor may condense onto frozen droplets of ocean water as they are transported to the surface (Schmidt et al., 2008). This means that the apparent concentration of chloride would be lower than the actual concentration in the ocean, so strictly speaking the reported range should be treated as a lower limit (Postberg et al., 2009). Second, the laser experiments may not provide a completely accurate representation of the impact ionization process at Enceladus. Postberg et al. (2009) argue otherwise, but it is impossible to know for sure unless calibration curves for impact and laser ionization of salt-bearing solutions/ices can be compared. Based on this discussion, the range 0.05-0.2 molal Cl$^-$ should be regarded as a useful but not rigid constraint on the geochemistry of Enceladus' ocean.

An alternative approach is to estimate the chloride concentration from the size of the ocean (Glein & Shock, 2010). In aqueous geochemistry, chloride is classified as a conservative species because it is very stable in liquid water, and is generally not incorporated into minerals except under arid conditions (Drever, 1997). This is useful behavior because the concentration of chloride in solution simply depends on the total amount of chloride and the mass of water in the environment of interest. Thus, the next step is to estimate how much chloride there is on Enceladus, and the mass of its ocean.

The total amount of chloride can be estimated from the mass of rock, as Enceladus most likely accreted chloride as a constituent in rocky material (Sharp & Draper, 2013; Bockelée-Morvan et al., 2014). Iess et al. (2014) recently determined the gravity field of Enceladus, and found the mean moment of inertia to be consistent with a differentiated interior having a rocky core of density 2400 kg m$^{-3}$ and radius 192 km. This implies a mass of rock of $7.1 \times 10^{19}$ kg. It is notable that the modeled core density is in close agreement with the grain density of the CI chondrite Orgueil (2420±60 kg m$^{-3}$; Macke et al., 2011). This suggests that CI chondrites can be a reasonable analogue for the bulk chemistry of Enceladus' core. If Enceladus accreted rocks with a chloride content similar to CI chondrites (700 ppm; Lodders, 2003), then the total amount of chloride on Enceladus would be $1.4 \times 10^{18}$ mol. All or at least most of this chloride can be expected to be dissolved in the ocean because the relatively low core density implies the presence of abundant hydrated silicates that are indicative of extensive water-rock interaction (see Section 4.3), and chloride would be readily leached into the water during such processes owing to its conservative nature (Zolotov, 2007; 2012). A total extraction scenario represents a plausible endmember, but it is possible that some of the chloride is not in the ocean; and could have been lost to Saturn's E ring, or trapped in the icy crust of Enceladus after being delivered there inside plume particles (Postberg et al., 2009; 2011).

The next step is to estimate the mass of ocean water. Glein & Shock (2010) presented an idealized model where the ocean is treated as a cap of uniform thickness that resides on top of the rocky core, and is centered on the geologically active south pole (Porco et al., 2006). In this model, the ocean can be a global or regional body of water (Collins & Goodman, 2007; Tobie et al., 2008; Běhounková et al., 2012) depending on how far north it extends in latitude (LAT; northern latitudes are positive, whereas southern latitudes are negative). While this model is a simplification (e.g., a more realistic ocean may be deepest beneath the south pole, becoming shallower at more northern latitudes; Iess et al., 2014), it allows a first-order estimate of the ocean's volume to be computed readily using a convenient analytical expression (Glein & Shock, 2010)

$$V_{ocean} = (2\pi/3)\left[r_{ocean}^3 - r_{core}^3\right]\left[1 - \cos(\text{LAT}+90°)\right], \qquad (3)$$

where the radius of the core is taken to be 192 km (Iess et al., 2014), and the radius of the ocean is related to Enceladus' mean radius ($R$ = 252 km) and the thickness of ice above the ocean by $r_{ocean} = R - h_{ice}$. The mass of ocean water can be obtained by multiplying its volume and density, the latter assumed to be similar to that of pure liquid water (1000 kg m$^{-3}$) as the plume particles are not very salty (Postberg et al., 2009; 2011).

The expected molal concentration of Cl$^-$ in Enceladus' ocean can now be calculated as a function of its latitudinal extent and the thickness of overlying ice

$$m_{Cl^-} = \left(6.7 \times 10^5\right)\left[\left(R - h_{ice}\right)^3 - r_{core}^3\right]^{-1}\left[1 - \cos(\text{LAT}+90)\right]^{-1}, \tag{4}$$

where the radii and ice thickness are expressed in units of kilometers. As expected, Fig. 1 shows that the Cl⁻ concentration is inversely related to the size of the ocean (Glein & Shock, 2010). The concentration is predicted to increase as the ice above the ocean thickens, and as the ocean becomes localized to more southern latitudes.

[Figure 1]

Plausible limits on the structure of the ocean can be derived from geophysical and geological arguments. From an analysis of the gravity and topography of Enceladus, Iess et al. (2014) inferred that isostatic compensation is taking place at the base of a layer of ice floating on Enceladus' ocean, and they estimated that the ice layer should be 30-40 km thick. This is one constraint. The latitudinal extent of the ocean is presently unknown (Iess et al., 2014), but a reasonable estimate is to assume that it underlies at least the active south polar region (Porco et al., 2006; Iess et al., 2014), and it may extend as far north as the equator (Tobie et al., 2008; Běhounková et al., 2012; Spencer & Nimmo, 2013). So, a second constraint is −50° < LAT < 0°. An ocean that underlies a significant portion of the northern hemisphere or a global ocean seems unlikely (but not impossible) because the heavily cratered north has been less geologically active than the south (Spencer et al., 2009), and their morphological differences can be attributed to the absence of a subsurface ocean and associated tidal heating in the north (Tobie et al., 2008). Another argument against an extensive ocean (LAT > 0°) is the difficulty in preventing a large ocean from freezing over geologic timescales (Roberts & Nimmo, 2008). Combining both structural constraints, the geophysical approach suggests that Enceladus' ocean can be expected to contain 0.2-1.2 molal $Cl^-$ (Fig. 1). Like for the geochemical approach, this should be regarded as a useful but not rigid constraint.

Altogether, the geochemical approach indicates that the ocean should contain 0.05-0.2 molal $Cl^-$, while the geophysical approach predicts 0.2-1.2 molal $Cl^-$ (Fig. 1). Therefore, the most likely value may be 0.2 molal $Cl^-$. In recognition of possible uncertainties in the estimates (see above), however, the entire range (0.05-1.2 molal $Cl^-$) should be considered. This range is significantly lower than the saturation concentration (~6 molal) of the mineral hydrohalite (Fig. 1), which supports the expectation of conservative behavior for chloride inside Enceladus (Glein & Shock, 2010).

The concentration of dissolved inorganic carbon (DIC = $\sum HCO_3^- + \sum CO_3^{-2}$) in the ocean can be estimated from the DIC/Cl molar ratio in the plume. Postberg et al. (2009) found that mass spectra of salt-rich plume particles can be reproduced well from laboratory solutions containing 2-5 times more $Cl^-$ than DIC. As discussed above, there is the question of whether these experiments provide a realistic simulation of the eruption and detection processes at Enceladus. On the other hand, the ratio should be more robust than the absolute concentration owing to canceling effects. As an example, condensation of steam on plume particles (Schmidt et al., 2008) would decrease the DIC concentration of the particles, but the concentration of $Cl^-$ would decrease correspondingly, such that the DIC/Cl ratio would be unchanged. Likewise, the detector response at Enceladus could differ from that in the laboratory, but the change in response for chloride and carbonate species may be similar, such that the ratio at Enceladus can be derived using laboratory data. However, this cancellation may not be perfect, so a conservative DIC/Cl ratio of 0.1-1 can be adopted to (plausibly) encompass the actual value. This would lead to a possible range of 0.005-1.2 molal for the DIC concentration, with a preferred range of 0.04-0.1 molal for the nominal case of 0.2 molal Cl⁻.

## 3. THERMODYNAMIC MODELS

### 3.1. Textbook speciation

*3.1.1. Method*

The simplest starting point for elucidating the geochemistry of Enceladus' ocean is to compute chemical equilibrium between aqueous $CO_2$, $HCO_3^-$, and $CO_3^{-2}$ at the plume-ocean interface. This textbook case of carbonate equilibria (Garrels & Christ, 1965; Drever, 1997) can be represented by the following chemical equations

$$CO_2(aq) + H_2O(l) \leftrightarrow HCO_3^-(aq) + H^+(aq), \tag{5}$$

$$HCO_3^-(aq) \leftrightarrow CO_3^{-2}(aq) + H^+(aq), \tag{6}$$

and the law of mass action leads to the corresponding equilibrium constant expressions

$$K_5 \approx m_{HCO_3^-} a_{H^+} / a_{CO_2} = 10^{-6.56} \text{ (at 0°C, 1 bar)}, \tag{7}$$

$$K_6 \approx m_{CO_3^{-2}} a_{H^+} / m_{HCO_3^-} = 10^{-10.62} \text{ (at 0°C, 1 bar)}, \tag{8}$$

where $m_i$ stands for the molal concentration of species $i$, and the equations are only approximately correct because the water is assumed to be pure (i.e., $a_{H2O} = 1$) and ideal behavior is also assumed (i.e., activity = molality). These are the conventional assumptions (Stumm & Morgan, 1996). There are four solutes in this model, so four constraints are required to calculate equilibrium. They are Eqs. (7) and (8), the activity of $CO_2$, and the molality of dissolved inorganic carbon in Enceladus' ocean

$$m_{DIC} = m_{HCO_3^-} + m_{CO_3^{-2}}. \tag{9}$$

Equations (7-9) can be combined and the resulting quadratic can be solved to obtain the activity of $H^+$ as a function of the activity of $CO_2$ and the molality of DIC

$$a_{H^+} \approx 0.5 \left( K_5 a_{CO_2} + \sqrt{K_5^2 a_{CO_2}^2 + 4 K_5 K_6 m_{DIC} a_{CO_2}} \right) m_{DIC}^{-1}. \tag{10}$$

Lastly, the pH of the ocean can be calculated from the definition (pH = −log[$a_{H+}$]).

*3.1.2. Results*

Figure 2 shows the calculated ocean pH for this geochemical model. It can be seen that the pH is predicted to decrease with increasing activity of $CO_2$, whereas it is directly related to the DIC concentration. In particular, the pH is a linear function of log($a_{CO2}$) at the most negative values of the latter quantity. At these conditions, $HCO_3^-$ is not important to the speciation, where essentially all of the inorganic carbon exists as $CO_3^{-2}$ because the pH is so high. In this regime, the pH can simply be written as

$$\text{pH} \approx 0.5\log(m_{DIC}) - 0.5\log(K_5) - 0.5\log(K_6) - 0.5\log(a_{CO_2}). \tag{11}$$

[Figure 2]

The pH of Enceladus' ocean can be approximated by adopting the constraints on the activity of $CO_2$ (from Section 2.2) and molality of DIC (from Section 2.3) in the ocean. Based on what is believed to be the most probable range in the DIC concentration (0.04-0.1 molal), this model would suggest that the ocean has a pH of 12.6±1.1 (Fig. 2). If it is assumed that the DIC concentration is known to less certainty (0.005-1.2 molal), then the uncertainty in the computed pH would increase somewhat (pH = 12.6±1.6; Fig. 2). It must be emphasized that these are illustrative values because the present model provides a simplified description of the carbonate system (see Section 3.2).

Nevertheless, the model allows several key generalizations to be made. First, the pH of the ocean should be much higher than neutral (Fig. 2), so the ocean should be strongly basic. For the ocean to instead have a circumneutral pH, Enceladus' plume gas would need to be composed almost entirely of $CO_2$ (Eq. 2), which is not the case (Waite et al., 2006; 2009; 2011; 2013). The geochemical factor that is most indicative of the high pH is the low activity of $CO_2$. This quantity is also responsible for most of the uncertainty in the derived pH (Fig. 2), which can be traced back to uncertainty in the temperature of the tiger stripes (Eq. 2; Goguen et al., 2013). Also, the inferred high ocean pH implies that the carbonate-bearing cluster ($Na_3CO_3^+$) identified by Postberg et al. (2009) in the plume particles was derived from aqueous carbonate, as opposed to the bicarbonate ion. The conclusion is that Enceladus' ocean can be classified geochemically as a Na-Cl-$CO_3$-type water (Deocampo & Jones, 2014), similar to but also different from Na-Cl-Mg-$SO_4$ seawater on Earth (Garrels & Thompson, 1962). It may seem remarkable that such a high pH solution can generate a plume containing appreciable amounts of $CO_2$ (Waite et al., 2011; 2013), which is a weakly acidic gas; but irreversible removal of $CO_2$ from the system by degassing and loss to space continuously drives the following reaction forward

$$CO_3^{-2}(aq) + 2H^+(aq) \rightarrow CO_2(g) + H_2O(l). \tag{12}$$

The ocean-plume system is thus demonstrating Le Chatelier's principle on a large scale. This equation also shows that $CO_2$ degassing can raise the pH, which occurs commonly when carbonate-saturated groundwaters that are in equilibrium with elevated fugacities of $CO_2$ emerge as springs at the Earth's surface (Hammer et al., 2005). Thus, $CO_2$ degassing may be a key process that has led to a high pH ocean on Enceladus, although it may not be as important as water-rock interactions and mineral buffering (see Sections 3.2.2.4 and 4).

## 3.2. Complete speciation

*3.2.1. Method*

The pH of Enceladus' ocean can be determined as accurately as (presently) possible by making a chemical model that accounts for equilibrium among all aqueous species (Garrels & Thompson, 1962). This is done here by performing thermodynamic calculations in the $Na_2O$-$HCl$-$CO_2$-$H_2O$ system at 0°C and 1 bar using the SpecE8 program, which is part of the Geochemist's Workbench® 6 software package (Bethke, 2008). This program calculates the activities and concentrations of aqueous species that satisfy all of the mass action (equilibrium constant) equations in the system, as well as conservation of mass and charge balance (electroneutrality). The present system contains the following 13 species: $Cl^-$, $CO_2(aq)$, $CO_3^{-2}$, $H^+$, $HCl$, $HCO_3^-$, $H_2O$, $Na^+$, $NaCl$, $NaCO_3^-$, $NaHCO_3$, $NaOH$, and $OH^-$. Equilibrium constants are taken from the standard thermo.dat database that is included with the Geochemist's Workbench®, and was compiled by the geochemical modeling group at Lawrence Livermore National Laboratory. This database is largely based on the SUPCRT database (Johnson et al., 1992). The system is treated as a non-ideal solution, and activity coefficients are computed using the B-dot equation of Helgeson (1969), which is an empirically extended form of the classic Debye-Hückel equation. The B-dot equation is thought to be reasonably accurate for ionic strengths up to ~1 molal (Bethke, 2008). For simplicity, the activity coefficients of neutral species are set to unity (Anderson, 2005). Oxidation-reduction reactions (Glein et al., 2008) and reactions involving minerals (Zolotov, 2007) do not need to be considered to calculate the pH of the ocean because the analytical data from Enceladus (see Section 2) are sufficient in this chemical system.

To solve for the equilibrium distribution, SpecE8 requires a certain number of input constraints, consistent with the phase rule. The set of components in the thermodynamic model is known as the basis, and the individual components are called basis species. The basis provides a mathematical representation of the bulk composition of the system of interest. For the present system, the default basis in SpecE8 contains $Cl^-$, $H^+$, $HCO_3^-$, $H_2O$ (= 1 kg), and $Na^+$. Components $Cl^-$ and $HCO_3^-$ correspond to total chloride and dissolved inorganic carbon, respectively. Analytical data from Enceladus (see Section 2.3) can be used to assign concentration values to these components. Because the activity of $CO_2$ in Enceladus' ocean has been constrained (see Section 2.2), it is convenient to swap $CO_2(aq)$ for basis species $H^+$. Because $Na^+$ is inferred to be the dominant cation in Enceladus' ocean (Postberg et al., 2009), its concentration is treated as an adjustable parameter to balance the negative charges of $Cl^-$, $CO_3^{-2}$, $HCO_3^-$, and $OH^-$. With these input constraints, the program can compute the pH and speciation for this model of Enceladus' ocean by solving numerically the equations of equilibrium using the Newton-Raphson method, with a convergence criterion of $5\times10^{-11}$ (Bethke, 2008).

*3.2.2. Results*

3.2.2.1. The pH

Figure 3 shows the calculated ocean pH at low (A), nominal (B), and high (C) $Cl^-$ concentrations for the complete speciation model. The pH curves in Fig. 3 bear a striking resemblance to those in Fig. 2, which makes sense because the simplified model in Section 3.1 captures the essentials of the carbonate geochemistry that are elaborated in the more sophisticated and realistic model in this section. The models differ, however, in their quantitative details; and the present model gives pH values that are slightly lower than those from the previous model ($\Delta pH \approx 0.2\text{-}0.7$). It is found, perhaps unsurprisingly, that the pH difference depends on composition; and becomes larger with increasing concentrations of $Cl^-$ and DIC, because the second-order effects of solution non-ideality and complex formation (e.g., $Na^+ + CO_3^{-2} \leftrightarrow NaCO_3^-$) become more important. Nevertheless, it can be concluded that the analytical model (Eq. 10) provides a reasonable approximation to the pH, and can be a useful tool for performing a quick calculation of planetary data that may have error bars larger than that of the model output.

[Figure 3]

The nominal case indicates that Enceladus' ocean should have a pH of 12.2±1.1 (Fig. 3B). However, a comprehensive analysis of the constrained compositional space suggests that the pH could be as low as 10.8 (Fig. 3A) or as high as 13.5 (Fig. 3C). This leads to a conservative estimate of the pH = 12.2±1.4. Overall, the more accurate speciation model reinforces the earlier conclusion that Enceladus' ocean should be strongly basic. Thus, the Enceladus ocean is clearly different from terrestrial seawater (pH ≈ 8.1; Garrels & Thompson, 1962), which is in chemical communication with a much larger reservoir of $CO_2$ gas. The inferred pH is similar to those (~11; ~12) predicted by Zolotov (2007) and Zolensky et al. (1989), respectively, from reaction path models of aqueous alteration of chondritic rock at 0°C and 100 bar. However, it is considerably higher than the value (~6-7) estimated by Marion et al. (2012) from a thermodynamic analysis similar to the present one. Yet, the Marion et al. (2012) model is based on an assumed partial pressure of $CO_2$ in the ocean that is thought to be unrealistically high (see Section 2.2), which led to the weakly acidic pH. The present pH is also somewhat higher than that (8.5-9) suggested by Postberg et al. (2009) from their interpretation of *Cassini* CDA data for reasons that have yet to be identified (see Section 1). Our value is consistent with both the CDA and INMS data, although the high pH is determined primarily by the abundance of $CO_2$ gas in the plume from INMS, and by our model of how its partial pressure changes going from the plume/tiger stripes to the ocean (i.e., Eq. 1). The present pH may be more robust since it is more consistent with what is expected for water-chondrite equilibrium at 0°C (Zolensky et al., 1989; Zolotov, 2007). This can also be taken as an argument that questions the pH (8-10) proposed recently by Hsu et al. (2014) from their hydrothermal model of nanosilica synthesis (Sekine et al., 2014). Therefore, alternative mechanisms of nanosilica synthesis (e.g., flash freezing of $SiO_2$-bearing ocean water) should be given more attention (Zolotov & Postberg, 2014).

3.2.2.2. Speciation

Figure 4 shows the chemical speciation that corresponds to the results in Fig. 3. $Na^+$ is computed to be the most abundant species in Enceladus' ocean, consistent with its ubiquity in the plume particles (Postberg et al., 2009; 2011); and $Cl^-$ should be the most abundant anion in the ocean. In general, the model implies that the ocean is dominated by an NaCl component, like terrestrial seawater (Garrels & Thompson, 1962). This attests to the great thermodynamic stability of these species in liquid water at low temperatures (Shock et al., 1997); and to the relatively high cosmic abundances of these elements in the rocky materials that were accreted by planetary bodies (Lodders, 2003). Indeed, this intuitive explanation is supported by the presence of NaCl-rich fluid inclusions in the Monahans meteorite (Zolensky et al., 1999), and chloride salts on the surface of Mars (Osterloo et al., 2010); and by the numerical modeling of Zolotov (2007), which predicts that NaCl should be the most abundant solute component in the Enceladus ocean.

[Figure 4]

Unlike seawater (<1 mmolal; Garrels & Thompson, 1962; Bethke, 2008), however, the Enceladus ocean is expected to contain significant amounts of the carbonate ion, where the carbonate speciation in the constrained range of $a_{CO2}$ depends on the concentrations of $Cl^-$ and DIC (Fig. 4). At lower concentrations of $Cl^-$ and DIC, $CO_3^{-2}$ is inferred to be the dominant form of DIC (Fig. 4A-D), while $NaCO_3^-$ would be more abundant at higher $Cl^-$ and DIC concentrations (Fig. 4E, F) because formation of the complex is driven by higher concentrations of $Na^+$ at these conditions. The source of the relatively high carbonate concentration may be buffering by a relatively soluble carbonate mineral, such as gaylussite [$Na_2Ca(CO_3)_2 \cdot 5H_2O \leftrightarrow CaCO_3 + 5H_2O + 2Na^+ + CO_3^{-2}$], which is found in soda lakes (Bischoff et al., 1991). Carbonate minerals presumably formed when $CO_2$-bearing ice (Ootsubo et al., 2012) melted and reacted with silicate minerals, perhaps during water-rock differentiation (Schubert et al., 2007). Mineralogical and isotopic evidence suggest that such chemistry also occurred on the parent bodies of numerous carbonaceous chondrites (Grady et al., 2002).

The carbonate-rich composition draws attention to terrestrial soda lakes (Garrels & Mackenzie, 1967; Jones et al., 1977; Eugster & Hardie, 1978; Bischoff et al., 1993; Kempe & Kazmierczak, 2002) as possessing an analogous geochemical feature to the Enceladus ocean. Thus, it would not be improper to call Enceladus' ocean a "soda ocean", and we suggest that geochemical and microbiological studies of soda lakes on Earth (e.g., Mono Lake in the Great

Basin or Lake Magadi in the East African Rift Valley) may provide useful insights into the geochemistry and possible biology of Enceladus' ocean (Sorokin et al., 2014). However, it should be noted that the analogy is not perfect as terrestrial soda lakes (pH ≈ 9-10.5) are generally not quite as basic as the Enceladus ocean (pH = 12.2±1.4). While Antarctic lakes can be seen as suitable physical analogues to Enceladus' ocean, they are apparently less geochemically relevant because they have a much lower pH (~6-8; Murray et al., 2012; Christner et al., 2014) than the ocean.

The relatively high carbonate concentration should limit (via the common ion effect) the oceanic concentrations of many metals (particularly Ca, Mn, Sr, Ba) that can be incorporated into carbonate minerals, as exemplified by the following equilibrium

$$CaCO_3(Calcite) \leftrightarrow Ca^{+2}(aq) + CO_3^{-2}(aq). \tag{13}$$

This example implies that the activity of $Ca^{+2}$ in Enceladus' ocean (Johnson et al., 1992) would be $\sim 4\times10^{-7}$ for a representative $CO_3^{-2}$ activity of 0.01, and the corresponding concentration of $Ca^{+2}$ would be of order μmolal, which is several orders of magnitude lower than in seawater (~6000 μm; Garrels & Thompson, 1962; Bethke, 2008).

Because the pH of Enceladus' ocean appears to be so high (Fig. 3), bicarbonate species should be only minor constituents of the ocean (Fig. 4). In contrast, the hydroxide ion is expected to be one of the more abundant species in the ocean owing to the inferred high ocean pH (i.e., low activity of $H^+$) that shifts the water ionization equilibrium to the right ($H_2O \leftrightarrow H^+ + OH^-$). This means that base-catalyzed reactions may be important to the synthesis (Proskurowski et al., 2008; Lang et al., 2010), transformation (Cody et al., 2011), and degradation (e.g., hydrolysis of nitriles, esters, peptides, polysaccharides, nucleic acids) of potential organic compounds (Waite et al., 2009; 2011; 2013) in the ocean. Base catalysis may allow various organics to reach equilibrium with respect to water addition. Therefore, anomalously high (super-equilibrium) concentrations of water soluble but hydrolytically labile compounds (e.g., urea, acetyl thioesters) in the plume would imply active production to counteract hydrolysis, strongly suggesting dynamic organic or biochemistry inside Enceladus.

The inferred high ocean pH of ~12 can be expected to lead to trace levels of other metals (e.g., Mg, Fe, Ni, Zn) in the Enceladus ocean. This is because these metals can be sequestered in phyllosilicate or hydroxide minerals in the rocky core (see Section 4.1; Zolotov, 2007; Iess et al., 2014; Neubeck et al., 2014; Sekine et al., 2014), and their release into ocean water requires acid hydrolysis, as illustrated in the example below

$$2Mg_3Si_2O_5(OH)_4(\text{Chrysotile}) + 6H^+(aq) \leftrightarrow Mg_3Si_4O_{10}(OH)_2(\text{Talc}) + 6H_2O(l) + 3Mg^{+2}(aq), \qquad (14)$$

which suggests that the equilibrium concentration of $Mg^{+2}$ in the ocean (Johnson et al., 1992) should be very low (~10-100 nmolal), which is many orders of magnitude lower than in seawater (~40 mm; Garrels & Thompson, 1962; Bethke, 2008).

To help guide future data analysis, experimentation, modeling, and mission planning, Table 1 provides a detailed summary of what are believed to be the best speciation models of Enceladus' ocean, based on current chemical information.

[Table 1]

3.2.2.3. Organics

The inferred ocean pH also has major consequences for the stability of organic compounds with acidic and/or basic functional groups (Schulte & Shock, 2004). At 0°C, 1 bar, and pH = 12; carboxylic acids ($pK_a \approx 5$) would exist almost exclusively as the deprotonated carboxylate anions (e.g., acetate), but an appreciable fraction (~10%) may be present as Na-complexes (Shock & Koretsky, 1993). Conversely, the fraction of undissociated carboxylic acids (e.g., acetic acid) should be very small because the pH would be many units above the $pK_a$ (Shock, 1995). Amines ($pK_a \approx 11$), meanwhile, would exhibit the opposite behavior in Enceladus' ocean (Shock et al., 1997). Most amine molecules would exist in the neutral form (e.g., $CH_3NH_2$), although a non-negligible fraction (~10%) would be present as the protonated aminium cations (e.g., $CH_3NH_3^+$).

These effects are important to current and future searches for ocean-derived organic compounds in Enceladus' plume (Waite et al., 2009), and their astrobiological implications (McKay et al., 2008). Ionized organics behave similarly to the constituents of inorganic salts, so they would be non-volatile at the conditions of the ocean, which means that they should not be degassed from the ocean (Postberg et al., 2009). This is in contrast to neutrals that can go into the plume gas. Therefore, organic acids from the ocean would be frozen inside the plume particles (Postberg et al., 2008; 2009; 2011), while ocean-derived amines might be detectable in the plume gas (e.g., $NH_3$; Waite et al., 2009; 2011; 2013). The particular examples chosen here may be of special interest, as acetate is the most abundant water-soluble organic in carbonaceous chondrites (Yuen et al., 1984), while methylamine has been detected in comet Wild 2 samples (Glavin et al., 2008). Simple amino acids (such as those in meteorites and comets; Pizzarello et al., 2006; Elsila et al., 2009) with one carboxyl and one amino group (e.g., glycine) should exist predominately as aqueous monoanions, so they would be present in the plume particles and not in the gas phase if they are derived from Enceladus' ocean. *In situ* and sample return methods will need to be developed to non-destructively access and analyze astrobiologically interesting organic salts that would be embedded in ice particles.

3.2.2.4. Caveats

The simple model of the pH is informative, but it may have additional uncertainties that are not captured by the error bars. One process that would change the computed pH is if some of the $CO_2$ that is degassed from the ocean condenses in the tiger stripes during the geysering process, as is evidently the case (Brown et al., 2006). This would mean that the plume gas contains less $CO_2$ than it should in terms of the model. The activity of $CO_2$ in the ocean would then be higher than presently adopted, which would lead to a pH lower than currently calculated (Fig. 3). It is difficult to determine how large this effect might be, but we are inclined to regard it as minor because the pH is relatively insensitive to $a_{CO2}$. In fact, an increase in $a_{CO2}$ by an order-of-magnitude would decrease the pH by approximately half a unit (Eq. 11). So, ~99% of the ocean-derived $CO_2$ would need to be removed from the plume gas for the pH to decrease by one unit. $CO_2$ condensation would thus appear to be a second-order effect in the model, unless >99.99% of the $CO_2$ snows out, which may be unlikely because $CO_2$ is quite volatile.

In contrast, the actual pH would be higher than the model-derived value if dry ice or clathrate hydrates (Kieffer et al., 2006) in Enceladus' ice shell are contributing $CO_2$ to the plume. This is because less of the observed $CO_2$ gas would be from the ocean, which would imply a lower activity of $CO_2$ than presently adopted. While this hypothesis cannot be ruled out, it does not need to be invoked to account for the presence of $CO_2$ in the plume because carbonate in the ocean (Postberg et al., 2009) can already provide $CO_2$ (Eq. 12), although clathrates may be required to account for the abundances of some of the other plume gases (Bouquet et al., 2014). It is worth noting also that the logarithmic nature of the pH implies that very large contributions from dry sources would be required to shift it significantly.

On the other hand, the actual pH would be lower than the model-derived value if $CO_2$ degassing at the top of the ocean is a disequilibrium process. This could occur if droplets of ocean water freeze before they can release sufficient $CO_2$ (Eq. 12) to satisfy vapor-liquid equilibrium (i.e., Henry's law). In this model, the plume should contain more $CO_2$ gas than it actually does, which would imply a higher activity of $CO_2$ than presently adopted. Such disequilibrium at the ocean-vapor interface seems plausible because the plume particles contain salts (Postberg et al., 2009; 2011), and equilibrium freezing would exclude salts from the ice (Zolotov, 2007). Non-equilibrium (flash) freezing of ocean water can explain the observations (Spencer & Nimmo, 2013), and if the freezing process is sufficiently fast, the carbonate system may not achieve phase equilibrium during the degassing process. Nevertheless, the system would need to be quite far from equilibrium (e.g., <1% of reaction progress) to appreciably change the computed pH (Eq. 11; see above), so this may also be a second-order effect.

A final complication could arise if Enceladus' ocean is not well-mixed. In this case, we would be calculating the pH of the uppermost layer of the ocean, where the plume gases are being generated (Schmidt et al., 2008). If the rate of degassing is faster than mixing, a compositional gradient would develop. It may be unreasonable to expect complete mixing in the vertical direction because the ocean may be deep (~20-30 km; Iess et al., 2014), compared to the Mariana Trench (~11 km) for example. Degassing of $CO_2$ raises the pH in the degassing zone (Eq. 12), so the bulk ocean may have a pH lower than what we have estimated (Fig. 3). This discussion calls attention to the need for kinetic modeling to understand the consequences of the dynamical balance between freezing, degassing, and mixing (Vance & Goodman, 2009), to allow us to clarify how representative the plume's chemistry is to the bulk ocean. In addition, high-precision carbon and oxygen isotope measurements of $CO_2$ in the plume gas (Waite et al., 2009) and carbonate in the plume particles (Postberg et al., 2009) would allow us to determine whether these species were previously in equilibrium at the surface of the ocean.

Because three of the four secondary effects discussed above would lead to a decrease in the pH, it is tempting to assume that the derived value (12.2±1.4; see Section 3.2.2.1) is a slight overestimate. At present, it is not possible to quantify the systematic error because there are four free parameters, but a best estimate may be pH = 11.2±1.4 (see Section 4.3).

## 4. GEOCHEMICAL IMPLICATIONS

### 4.1. Serpentinization of chondritic rock

The present study represents a top-down approach of inferring the pH from observational data, while the study of Zolotov (2007) represents a bottom-up approach of predicting the pH from physiochemical modeling of expected water-rock interactions. The fact that these independent models are converging on a pH of ~11-12 can be taken as an argument that the ocean has this pH. Their consistency also suggests that the assumptions underlying these approaches are appropriate to Enceladus, and provides confidence that they can be used to accurately predict the geochemistry of subsurface oceans (Zolotov & Kargel, 2009) and the composition of associated plumes (Roth et al., 2014) on other icy bodies. Also, the general consistency of the pH from the theoretical model with the observationally-derived value can be taken as evidence that water-rock interactions are the primary controller of the pH of Enceladus' ocean (Zolotov, 2007).

Consistent with the mineralogy of aqueously altered chondrites (Brearley & Jones, 1998; Brearley, 2006), the relatively low density of Enceladus' rocky core (Iess et al., 2014), and experimental (Sekine et al., 2014) and theoretical (Zolotov, 2007) simulations, serpentinization (McCollom & Seewald, 2013) can be seen as the key process that led to a high pH ocean on Enceladus, as shown by the (idealized) reaction below

$$2Mg_2SiO_4(Forsterite) + H_2O(l) + 2H^+(aq) \rightarrow Mg_3Si_2O_5(OH)_4(Chrysotile) + Mg^{+2}(aq), \quad (15)$$

which consumes protons and thus raises the pH. In fact, the deduced pH of the ocean (~11-12) is similar to those (~9-12.5) in hyperalkaline springs from numerous sites of low-temperature serpentinization on Earth (Barnes et al., 1967; 1978; Kelley et al., 2001; Mottl et al., 2003; Morrill et al., 2013; Szponar et al., 2013), which bolsters the expectation that serpentinization has occurred on Enceladus (Vance et al., 2007; Malamud & Prialnik, 2013). Thus, serpentinizing systems on Earth, such as the Semail ophiolite in Oman and the Lost City hydrothermal field in the mid-Atlantic Ocean, can be regarded as reasonable geochemical and perhaps biological analogues of the water-rock (i.e., ocean-core) system on Enceladus (see Section 5.2). This discussion echoes the emerging theme of serpentinization as the most important alteration reaction in the Solar System (Brearley, 2006; Schulte et al., 2006; Ehlmann et al., 2010), which suggests that subsurface oceans on other worlds in the outer Solar System (Hussmann et al., 2006; Castillo-Rogez & McCord, 2010; Robuchon & Nimmo, 2011; Hammond et al., 2013) would also be alkaline solutions (Zolotov, 2012).

**4.2. Hydrogen generation from iron oxidation**

An important consequence of serpentinization is the simultaneous generation of $H_2$ by aqueous oxidation of reduced iron-bearing minerals (Neal & Stanger, 1983; Sleep et al., 2004; Oze & Sharma, 2007; McCollom & Bach, 2009; Klein et al., 2013; Mayhew et al., 2013). The presence of ferric ($Fe^{+3}$) phases (e.g., magnetite, maghemite, cronstedtite) in aqueously altered chondrites (Brearley, 2006) demonstrates that this process can also occur in chondritic systems (Rosenberg et al., 2001; Alexander et al., 2010). The likely occurrence of serpentinization on Enceladus thus implies that $H_2$ has been produced inside Enceladus, where the dominant source of $H_2$ is suggested to be the oxidation of accreted metallic iron (which is abundant in primitive bodies and also unstable in the presence of liquid water at sub-kbar pressures; Brearley & Jones, 1998; Zolensky et al., 2006; Glein et al., 2008)

$$3Fe(Kamacite) + 4H_2O(l) \rightarrow Fe_3O_4(Magnetite) + 4H_2(aq). \qquad (16)$$

An estimate of the $H_2$-generating potential of the water-rock system on Enceladus can be made by assuming that Enceladus' rocky core ($7.1\times10^{19}$ kg; Iess et al., 2014) has a total Fe content similar to that in CI chondrites (~18%), and ~35% of the Fe was accreted as metal, with the remainder as troilite (FeS) and ferrous ($Fe^{+2}$) silicates (Lodders, 2003). This model could produce $1.1\times10^{20}$ mol of $H_2$ via the stoichiometry of Eq. (16). A total water (liquid+ice) mass of $3.7 \times 10^{19}$ kg (Iess et al., 2014) then leads to a bulk concentration of ~3 mol of $H_2$ per kg of water (cf. Lost City, ~0.01 molal $H_2$; Proskurowski et al., 2006). This high value illustrates that the water-rock system has the potential to make enormous quantities of $H_2$, as accreted rocks would transfer reducing potential from the $H_2$-rich solar nebula (Prinn & Fegley, 1989) to Enceladus' interior. The formation of $H_2$ inside Enceladus would provide a favorable environment for abiotic synthesis of organic molecules (Shock & Canovas, 2010; McCollom, 2013) that may have led to an origin of life (Russell et al., 2010), or for $H_2$-utilizing microorganisms (McKay et al., 2008; see Section 5). The molecular and isotopic composition of organic compounds in the plume (Waite et al., 2009) may thus contain evidence (e.g., light $\delta D$) of $H_2$-enabled synthesis in the interior (McKay et al., 2012).

However, the actual (e.g., steady-state) concentration of $H_2$ in the ocean could be considerably lower than the theoretical maximum if plume formation (Waite et al., 2011; Brockwell et al., 2014) or other loss processes are depleting the ocean in $H_2$, or if not all of the reduced Fe minerals have been oxidized (see Section 4.3). As an example, the above endmember scenario implies a mean production rate of $H_2$ of 775 mol sec$^{-1}$ over the history of the Solar System ($4.5 \times 10^9$ y). A flux of 200 kg (11,100 mol) of water vapor per sec in Enceladus' plume (Hansen et al., 2011) would then lead to a predicted native $H_2/H_2O$ ratio on the order of 10% in the plume gas, which should be detectable to *Cassini* INMS (Brockwell et al., 2014). Interestingly, recent work shows that the measured abundance of $H_2$ agrees well with the predicted value, and the abundance of native $H_2$ could be as high as ~15% relative to $H_2O$ (Waite et al., 2013). The presence of native $H_2$ in the plume would have profound implications for the oxidation state (e.g., $H_2$ fugacity) of the ocean and underlying rocks (Glein et al., 2008).

**4.3. Extensive alteration and buffering of the ocean**

At present, it is unknown whether serpentinization is still occurring on Enceladus. The anhydrous minerals (Brearley & Jones, 1998; Zolensky et al., 2006) that were accreted by Enceladus may have been completely serpentinized during past aqueous alteration (Vance et al., 2007; Neveu et al., 2014), so Enceladus may lack feedstock for contemporary serpentinization. This possibility would be consistent with Enceladus having a relatively low core density (2400 kg m$^{-3}$; Iess et al., 2014) that is similar to those of many phyllosilicate minerals in carbonaceous chondrites (Brearley & Jones, 1998), such as chrysotile serpentine (2500 kg m$^{-3}$). An anhydrous core can be expected to have a density similar to that of Jupiter's dehydrated moon, Io (3500 kg m$^{-3}$; Schubert et al., 2007). It would appear, then, that the most reasonable interpretation is that at least most of the rocky core was serpentinized in the past, although we cannot exclude the possibility of small amounts of unaltered rock in the deepest parts of Enceladus' core, which may not have a perceptible effect on the gravimetric data (Iess et al., 2014). If endogenic $H_2$ is present in the plume (Brockwell et al., 2014), this would provide compelling evidence for current aqueous alteration processes on Enceladus (Waite et al., 2013). Assuming isotopic equilibrium (Neal & Stanger, 1983), we predict that $H_2$ from the ocean would have a D/H ratio ~4.5 times smaller than that in water (Horibe & Craig, 1995).

The inorganic chemistry of today's ocean is likely to be controlled by equilibria (Sillén, 1961; 1967) between the products of serpentinization, such as phyllosilicates and carbonates, and the aqueous solution. Indeed, equilibrium geochemistry (Zolensky et al., 1989; Zolotov, 2007) can explain why Enceladus' ocean (evidently) has a pH of ~11-12. Thermodynamic calculations to be reported in a forthcoming publication indicate that an ocean floor assemblage of chrysotile-talc-dolomite-calcite-gaylussite in the presence of 0.2 molal Cl⁻ would buffer the pH = 11.4, $\log(a_{CO_2}) = -7.5$, and $m_{DIC} = 0.07$ molal at 0°C and 1 bar. These predicted values are all intriguingly consistent with the constraints derived from the observational data (see Section 2). Equilibrium between sediments and ocean water may occur on Enceladus but not on Earth (Broecker, 1971; Berner & Berner, 1996), because the hydrosphere on Enceladus may be less compositionally heterogeneous (no continental crust) and dynamically disturbed (no solar-driven water cycle) than that on Earth. In the absence of large rates of energy input (tidal heating at Enceladus is dwarfed by solar heating at Earth; Meyer & Wisdom, 2007) and mixing, Enceladus' ocean may equilibrate with many chemical elements in marine sediments over geologic time (apart from kinetically inhibited redox reactions of carbon, nitrogen, sulfur, and possibly iron; Zolotov & Shock, 2004).

## 4.4. Limited impact of oxidants from the surface

Besides being consistent with serpentinization, the inferred high ocean pH (~11-12; see Section 3.2.2.4) can rule out or restrict other geochemical processes. For example, strong oxidants ($H_2O_2$, $O_2$, $O_3$) are expected to form on the surface of Enceladus as a result of photolytic and radiolytic processing of water ice (Cooper et al., 2009). The delivery of these oxidants to the ocean (Hand et al., 2007) is an interesting possibility because they would affect the oxidation state of the system (Glein et al., 2008), and could support metabolisms that yield more chemical energy (Chyba, 2000; Catling et al., 2005). A less obvious side effect of oxidant delivery would be a decrease in the pH of the ocean, in response to sulfuric acid production (Pasek & Greenberg, 2012). The geochemistry of oxidant delivery to oceans on icy moons can be envisioned to parallel that of acid mine drainage on Earth (Drever, 1997), as exemplified by the following reaction

$$4FeS(\text{Troilite}) + 6H_2O(l) + 9O_2(aq) \rightarrow 4FeOOH(\text{Goethite}) + 8H^+(aq) + 4SO_4^{-2}(aq). \quad (17)$$

The combined clues that the ocean is highly alkaline (i.e., low activity of $H^+$), and sulfate-bearing clusters have not been reported to be present in the plume particles (Postberg et al., 2009; 2011) suggest that the oxidation of metal sulfides to sulfuric acid has not been a quantitatively significant process over the history of Enceladus, unlike on Europa perhaps (Hand et al., 2007; although see McKinnon & Zolensky, 2003 and Brown & Hand, 2013 for a different perspective). Indeed, this would also be thermodynamically consistent with the presence of reduced gases and organics in Enceladus' plume (Waite et al., 2009; 2011; 2013). Carbonate minerals on the ocean floor could maintain the high pH by neutralizing sulfuric acid, but sulfate would be produced; hence, this hypothesis requires an explanation for the (apparent) absence of sulfate salts in Enceladus' plume.

It may be surprising that exogenous oxidants and the associated production of sulfuric acid are apparently not more important to the geochemistry of Enceladus because water-derived oxidants should be produced continuously on the surface, as evidenced by their presence on some of Saturn's other icy satellites (Noll et al., 1997; Teolis et al., 2010; Tokar et al., 2012); and active resurfacing at the south pole (Porco et al., 2006) provides a mechanism for delivering them to the ocean (Hand et al., 2007). A potential solution to this puzzle is that the resurfacing mechanism (geysering) may be active infrequently (Spencer & Nimmo, 2013), which would limit the delivery of oxidants to the ocean, keeping the pH high and sulfate low. This episodic model is the simplest and perhaps most likely explanation. A more speculative possibility is that life could be reversing the effects of Eq. (17) via biologic sulfate reduction (Zolotov & Shock, 2003)

$$2FeOOH(Goethite) + 8H^+(aq) + 4SO_4^{-2}(aq) + 15H_2(aq) \rightarrow 2FeS_2(Pyrite) + 20H_2O(l), \quad (18)$$

which would provide an alternative explanation for why the ocean is not more acidic and richer in sulfate, but this mechanism requires the presence of organisms and reductants (e.g., $H_2$ from serpentinization or organic carbon; see Section 5.2) because the abiotic process is kinetically inhibited at low temperatures (Ohmoto & Lasaga, 1982).

**4.5. Absence of basaltic rock on the seafloor**

Another implication is that the suggested ocean pH (11.2±1.4) is so high that it argues against water-basalt reactions on Enceladus. Waters from terrestrial basalt aquifers are only mildly basic (pH ≈ 8-9; Wood & Low, 1986; Gislason & Eugster, 1987), mainly because basalts are poorer in the basic component MgO and enriched in non-basic $Al_2O_3$ and $SiO_2$ relative to peridotites and chondrites (Jarosewich, 1990). Hence, the high pH of Enceladus' ocean implies that the ocean floor should not have a basaltic (or more felsic) composition (Wetzel & Shock, 2000), unlike the Earth's oceanic crust and the surfaces of the other terrestrial planets (Basaltic Volcanism Study Project, 1981; McSween et al., 2006; Nittler et al., 2011). Otherwise, the pH would be lower (~8-9). What this means is that the ultrabasic core of Enceladus never got hot enough for the rocks to partially melt (i.e., to cross the wet solidus at ~1100-1200°C), which is necessary for the petrogenesis of basaltic rocks (Philpotts & Ague, 2009). This may not be terribly surprising because Enceladus is a small world, and radiogenic heating generally leads to higher temperatures inside bigger bodies (Turcotte & Schubert, 2002). A lack of basalt would also be consistent with the apparent absence of a dense metal inner core (Iess et al., 2014), whose formation would also require melting ($T > 900\text{-}1000°C$; Buono & Walker, 2011). This is in contrast to Jupiter's moon, Europa, which is thought to have a metallic core (Schubert et al., 2009) and probably basaltic rocks on its ocean floor (Zolotov & Kargel, 2009). The absence of silicate volcanism implies that Enceladus lacks a mechanism of recycling rock. As a result, accreted rocks can only be serpentinized once (see Section 4.3).

On the other hand, Enceladus is similar in size ($R = 252$ km) to the asteroid Vesta ($R = 263$ km), which experienced igneous differentiation (Russell et al., 2012). Why did basalt (i.e., eucrites) form on Vesta (McSween et al., 2013) but not on Enceladus? There seem to be two reasons. First, large amounts of water ice on Enceladus (Iess et al., 2014) can act as a substantial thermal buffer (i.e., the latent heat of fusion), while Vesta appears to lack significant ice based on its high bulk density (3456 kg m$^{-3}$; Russell et al., 2012). Second, Enceladus may have formed later than Vesta, after much of the potent but short-lived heat source $^{26}$Al had decayed (Jacobsen et al., 2008). The truth probably lies in a combination of these factors, which suggests that the inferred lack of a basaltic ocean floor and metal inner core can constrain models of the thermal evolution of Enceladus (Castillo-Rogez et al., 2007b; Schubert et al., 2007; Neveu et al., 2014) and origin of the Saturnian system (Castillo-Rogez et al., 2007a; Canup, 2010; Asphaug & Reufer, 2013). While igneous activity is unlikely on Enceladus, this does not preclude past or present hydrothermal activity (Matson et al., 2007; Glein et al., 2008; Sekine et al., 2014).

## 5. LIFE ON ENCELADUS

Given our improved understanding of the geochemistry of Enceladus' ocean, it is natural to conclude with a discussion of the possibility of life (McKay et al., 2014), which is a topic of intense scientific and societal interest, and a major motivator of space exploration. There are two key issues that determine if life exists in the Enceladus ocean (Hand et al., 2009): (1) the acquisition of life, and (2) a sustained production of key chemicals essential for life as we know it. The high alkalinity ocean on Enceladus (see Section 3.2.2.1) inferred from the observed carbonate geochemistry of the plume implies that serpentinization occurred (see Section 4.1) and could still be occurring on Enceladus (see Section 4.3). What are the implications of serpentinization (past or present) for Enceladus to acquire and sustain life as we know it? Recent considerations point to serpentinization as an important process in the origin of life on Earth. Moreover, microorganisms have been identified that grow in these environments at very alkaline pH, over a wide range of temperatures, in the absence or presence of $O_2$; and they employ metabolisms that are dependent on $H_2$. Below, we elaborate on these topics.

## 5.1. Serpentinization and the origin of life

We assume that it is improbable that Enceladus acquired life from Earth (Worth et al., 2013); if life exists on Enceladus, it presumably formed *de novo*. There are many models for how life formed on Earth (Fry, 2000), and most require a complex set of geo-settings including hydrothermal processes and possibly extraterrestrial input to provide some of the organic compounds needed for an origin of life (Stüeken et al., 2013). On Earth, there are only limited geological data from the Hadean and Archean Eons, but the available evidence indicates that the crust was significantly more ultramafic than at present, which would have led to more widespread serpentinization (Sleep et al., 2011). The 3.8 Ga Isua supercrustal belt of Western Greenland is an example of such an ancient environment that has features consistent with serpentinization and isotopic evidence for $H_2$-utilizing microbial communities (Nisbet and Sleep, 2001; Friend et al., 2002; Sleep et al., 2011). There are many properties of serpentinization that link this process to key steps in the origin of life, especially the generation of $H_2$ (see Section 4.2) to use as a reductant and energy source for biochemical processes. Serpentinization also promotes the formation of organic compounds (e.g., hydrocarbons, fatty acids) through Fischer-Tropsch-type reactions (Proskurowski et al., 2008; McCollom, 2013). Moreover, the polymerization of HCN is facile at high pH resulting in the synthesis of amino acids, purines, and pyrimidines (Ferris & Hagan, 1984; Kempe & Kazmierczak, 2011). Alkaline conditions are also highly conducive to the abiotic synthesis of carbohydrates, such as ribose, the backbone sugar of ribonucleic acid (RNA), via the formose reaction (Ricardo et al., 2004; Kim et al., 2011). All of this suggests that analogous abiotic syntheses should be possible on Enceladus as the plume contains many of the necessary ingredients, such as HCN, formaldehyde, and other organics (Waite et al., 2006; 2009; 2011; 2013).

A prominent model for the origin of life proposes that life on Earth began in a hydrothermal vent that was powered by serpentinization (Martin et al., 2014; Russell et al., 2014). An attractive feature of hydrothermal vents as hatcheries of life is the wide range of physical (e.g., temperature) and chemical (e.g., pH) gradients that arise from their dynamic properties (Baross & Hoffman, 1985; Martin et al., 2008). Serpentinization produces $H_2$-rich fluids, and studies of evolutionarily ancient organisms suggest that the earliest metabolic pathways were $H_2$-driven (Fuchs, 2011). The extant metabolic pathway that is considered to be the most promising model for the emergence of metabolism is the reductive acetyl-CoA pathway used by methane-producing archaea and acetate-producing bacteria (Russell & Martin, 2004). This pathway uses $H_2$, and is thought to be very ancient as the catalytic sites on some of the enzymes are transition metal-sulfur clusters that resemble certain minerals (Cody & Scott, 2007), and the proteins also appear to be phylogenetically and structurally ancient (Pereto et al., 1999; Berg et al., 2010).

The above considerations provide circumstantial evidence that life can arise in serpentinizing systems. The occurrence of serpentinization on Enceladus (see Section 4.1) therefore suggests that the geochemical environment on this icy world would be conducive to abiogenesis. Yet, serpentinization does not necessarily lead to life, given that carbonaceous chondrites do not appear to contain extraterrestrial organisms, despite the occurrence of serpentinization on many meteorite parent bodies (i.e., asteroids; Brearley, 2006). A simple hypothesis for the non-emergence of life in meteorites is that liquid water did not exist long enough (Abramov & Mojzsis, 2011). Some meteorites may have been on the road to life as they contain many of the molecules of life, such as amino acids, sugars, nucleobases, and metabolic intermediates (Pizzarello et al., 2006; Callahan et al., 2011; Cooper et al., 2011); but evidently they never got past the monomer stage. On Enceladus, at least a localized body of liquid water may have been maintained by tidal heating for perhaps the age of the Solar System (Tobie et al., 2008; Běhounková et al., 2012), so there may be more than enough time for life to originate. If serpentinization inevitably leads to life given sufficient time, then we should be able to test this by searching for evidence of past or present life in Enceladus' ocean. Alternatively, life may have never started on Enceladus if a lower pH, an atmosphere (Miller, 1953), plate tectonics, or access to sunlight or dry land (Powner et al., 2009) are necessary. Therefore, the search for evidence of prebiotic chemical evolution and life on Enceladus will give us an excellent opportunity to distinguish between essential vs. ancillary conditions and processes in understanding the origin of life as an outgrowth of geochemistry.

**5.2. Metabolic strategies of putative life**

If life did emerge *de novo* on Enceladus and if it mimics Earth life biochemically (Pace, 2001), what would be its characteristics given a lack of photosynthesis to fix carbon and generate $O_2$, and potentially a lack of other electron acceptors (i.e., oxidants) that are needed to produce the diversity of life as observed in Earth's oceans (Venter et al., 2004)? Are there terrestrial microorganisms that could grow in an Enceladus ocean that is highly alkaline (see Section 3.2.2.1) and anoxic (see Section 4.4) as a result of the geochemistry of serpentinization? Serpentinization environments on Earth can be regarded as reasonable analogues to the ocean-core geochemical system on Enceladus (although they are not perfect analogues because peridotites are not compositionally identical to chondrites; for example, chondrites contain more iron, nickel, and sulfur than peridotites; Jarosewich, 1990). By understanding the survival strategies of organisms in these environments, we can gain insights into how life could persist on Enceladus.

The discovery of the Lost City hydrothermal field in 2000 inspired research into understanding microbial life in serpentinization environments (Kelley et al., 2001; 2005), where the pH is high (~9-12.5) like the Enceladus ocean (~11-12; see Section 3.2.2.4). Serpentinization environments on Earth can be categorized as near-ambient (20-30°C) or higher-temperature (>80°C) systems, and they are found in both marine and continental settings. Lost City is the most famous example of a higher-temperature submarine system. Elevated temperatures can result from water-rock reactions that release heat when primary igneous minerals (e.g., olivine, pyroxene) are hydrated into serpentine and other alteration minerals. In terms of their microbiology, marine and continental serpentinization environments have both anaerobic and aerobic components (Brazelton et al., 2013; Schrenk et al., 2013). Here, we focus on the anaerobes since a large and consistent source of $O_2$ would not exist in the presumed absence of photosynthesis on Enceladus. In terrestrial serpentinites, $CH_4$ and $H_2$ are very important, consistent with the reducing conditions in ultramafic rocks at depth; indeed, $CH_4$ and $H_2$ metabolizing archaea and bacteria dominate these systems (Brazelton et al., 2006; 2011; Schrenk et al., 2013). At Lost City in particular, the microbial community is dominated by a single species of anaerobic $CH_4$ metabolizing archaea capable of both producing and oxidizing $CH_4$ (Brazelton et al., 2011). $H_2$ stimulates both of these metabolisms. In the presence of suitable electron acceptors such as $O_2$, $NO_3^-$, $Fe^{+3}$, or $SO_4^{-2}$, many species of bacteria and archaea can harvest chemical energy by oxidizing $CH_4$ or $H_2$ (Schrenk et al., 2013). The existence of life in these environments shows that the very alkaline pH is not an insurmountable barrier to survival, but high pH presents a challenge to Earth life because many biomolecules can be attacked by $OH^-$ (Westheimer, 1987). To prevent these deleterious effects, alkaliphilic organisms have evolved biochemical adaptations (e.g., $Na^+/H^+$ antiporters) to keep their cytosolic pH near neutral, despite the high external pH (Krulwich et al., 2011). We can assume that analogous life on Enceladus would also employ active transport as long as sufficient chemical energy is available to maintain the proton gradient.

We conclude that the most likely metabolic groups capable of growing in Enceladus' ocean would be anaerobic methanogens (reduce $CO_2$ to methane) and acetogens (reduce $CO_2$ to acetate; McCollom, 1999; Zolotov & Shock; 2004; Takai et al., 2013), although more exotic metabolisms may also be possible (Zolotov, 2010). $CO_2$ and carbonate are present in Enceladus' plume (Postberg et al., 2009; 2011; Waite et al., 2009; 2013), and our geochemical model indicates that the ocean is rich in carbonate (Table 1). Hence, the availability of inorganic carbon should not limit biological productivity, unlike in serpentinizing systems on Earth (Schrenk et al., 2013). A reducing agent that can sustain life on Enceladus would be $H_2$ produced by water-rock reactions (see Section 4.2). Based on the discussion of the origin of life (see Section 5.1), we would predict that the acetyl-CoA metabolic pathway or some modified form of this pathway would evolve at an early stage, to provide organic carbon and energy. If this were the case, then methanogens and acetogens may dominate by analogy to Earth life. Oligotrophic heterotrophs (organisms that grow on low levels of organic compounds) could also exist on Enceladus, given that various organics have been detected in Enceladus' plume (Waite et al., 2009). Fermentation may be a viable metabolic strategy for such organisms. In contrast, the oxidation of organic carbon may not be feasible owing to a limited availability of oxidants in the ocean (see Section 4.4).

Is there life on Enceladus? At present, we cannot unfortunately answer this question because we lack sufficient information. What we can say, however, is that based on our knowledge of life on Earth, the geochemistry of serpentinization on Enceladus would provide redox disequilibria between $H_2$, carbonate, $CH_4$, acetate, and perhaps other organic compounds that can support microorganisms. Aside from the origin of life, the biggest unknown that would be critical to life on Enceladus is the availability of $H_2$. Are there "fresh" anhydrous rocks deep in Enceladus' core, allowing $H_2$ to be produced today (Brockwell et al., 2014); or has the ocean-core system reached a state of complete chemical equilibrium (see Section 4.3)? If the latter, it is possible that life existed but died out because they ran out of food (Gaidos et al., 1999). This is an important issue to consider as we move forward in assessing the habitability of Enceladus (Shapiro & Schulze-Makuch, 2009; McKay et al., 2014). Ultimately, empirical observation will settle the life question, as the plume provides a very convenient means of searching for molecular and isotopic evidence of life (e.g., methane, acetate; McKay et al., 2008; Postberg et al., 2008; Waite et al., 2009), and possibly even frozen organisms (Tsou et al., 2012). Whatever the outcome, further exploration of Enceladus is likely to transform our understanding of the biological potential of icy worlds in the outer Solar System, and provide a more informed perspective on our place in a possible living Universe.

# ACKNOWLEDGEMENTS

This paper would not have come to pass without the unprecedented data that have been obtained by the *Cassini* mission, and it was inspired partly by the timeless work of Robert Garrels. C. R. G. is grateful to George Cody, Matthieu Galvez, Conel Alexander, Bjørn Mysen, and Frank Postberg for many interesting discussions and their encouragement; and he would also like to express his gratitude to the Carnegie Institution and NASA Astrobiology Institute for supporting his postdoctoral fellowship. C. R. G. and J. A. B. thank the Enceladus LIFE team (especially Ariel Anbar and Peter Tsou) for many pedagogical discussions about a sample return mission to Enceladus. J. A. B. acknowledges funding from NASA Astrobiology Institute Grant through Cooperative Agreement NNA04CC09A to the Geophysical Laboratory at the Carnegie Institution for Science. J. H. W. acknowledges funding from the Cassini Project (contract number NAS703001TONMO711123). This paper began as a presentation at the first Enceladus LIFE Workshop, and benefited from the stimulating atmosphere at the meeting. Subsequent discussions at the Europa Clipper Plume Advisory Session helped to refine our ideas. C. R. G. dedicates his contribution to this paper to S. I. G. and J. L. W. G.

Brazelton W. J., et al. (2013) Bacterial communities associated with subsurface geochemical processes in continental serpentinite springs. *Appl. Environ. Microbiol.* **79**, 3906-3916.

Brearley A. J. (2006) The action of water. In *Meteorites and the Early Solar System II* (eds. D. S. Lauretta, H. Y. McSween). Univ. of Arizona Press, Tucson, AZ. pp. 587-624.

Brearley A. J., Jones R. H. (1998) Chondritic meteorites. In *Planetary Materials* (ed. J. J. Papike). Mineral. Soc. of Am., Washington, DC. pp. 3-1—3-398.

Brockwell T. G., et al. (2014) Hydrogen in Enceladus' plume – Native or artifact? *Workshop on the Habitability of Icy Worlds*, Abstract #4022.

Broecker W. S. (1971) A kinetic model for the chemical composition of sea water. *Quaternary Res.* **1**, 188-207.

Brown M. E., Hand K. P. (2013) Salts and radiation products on the surface of Europa. *Astron. J.* **145**, 110, doi:10.1088/0004-6256/145/4/110.

Brown R. H., et al. (2006) Composition and physical properties of Enceladus' surface. *Science* **311**, 1425-1428.

Buono A. S., Walker D. (2011) The Fe-rich liquidus in the Fe-FeS system from 1 bar to 10 GPa. *Geochim. Cosmochim. Acta* **75**, 2072-2087.

Callahan M. P., et al. (2011) Carbonaceous meteorites contain a wide range of extraterrestrial nucleobases. *Proc. Natl. Acad. Sci. USA* **108**, 13,995-13,998.

Canup R. M. (2010) Origin of Saturn's rings and inner moons by mass removal from a lost Titan-sized satellite. *Nature* **468**, 943-946.

Castillo-Rogez J. C., McCord T. B. (2010) Ceres' evolution and present state constrained by shape data. *Icarus* **205**, 443-459.

Castillo-Rogez J. C., et al. (2007a) Iapetus' geophysics: Rotation rate, shape, and equatorial ridge. *Icarus* **190**, 179-202.

Castillo-Rogez J. C., et al. (2007b) The early history of Enceladus: Setting the scene for today's activity. *Lunar Planet. Sci. XXXVIII*, Abstract #2265.

Catling D. C., et al. (2005) Why $O_2$ is required by complex life on habitable planets and the concept of planetary "oxygenation time". *Astrobiology* **5**, 415-438.

Christner B. C., et al. (2014) A microbial ecosystem beneath the West Antarctic ice sheet. *Nature* **512**, 310-313.

Chyba C. F. (2000) Energy for microbial life on Europa. *Nature* **403**, 381-382.

Cody G. D., Scott J. H. (2007) The roots of metabolism. In *Planets and Life: The Emerging Science of Astrobiology* (eds. W. T. Sullivan, J. A. Baross). Cambridge Univ. Press, New York. pp. 174-186.

Cody G. D., et al. (2011) Establishing a molecular relationship between chondritic and cometary organic solids. *Proc. Natl. Acad. Sci. USA* **108**, 19,171-19,176.

Collins G. C., Goodman J. C. (2007) Enceladus' south polar sea. *Icarus* **189**, 72-82.
49

Spencer J. R., et al. (2009) Enceladus: An active cryovolcanic satellite. In *Saturn from Cassini-Huygens* (eds. M. K. Dougherty, L. W. Esposito, S. M. Krimigis). Springer, New York. pp. 683-724.

Spitale J. N., Porco C. C. (2007) Association of the jets of Enceladus with the warmest regions on its south-polar fractures. *Nature* **449**, 695-697.

Stüeken E. E., et al. (2013) Did life originate from a global chemical reactor? *Geobiology* **11**, 101-126.

Stumm W., Morgan J. J. (1996) *Aquatic Chemistry: Chemical Equilibria and Rates in Natural Waters*. John Wiley & Sons, New York.

Sverjensky D. A., Lee N. (2010) The great oxidation event and mineral diversification. *Elements* **6**, 31-36.

Szponar N., et al. (2013) Geochemistry of a continental site of serpentinization, the Tablelands Ophiolite, Gros Morne National Park: A Mars analogue. *Icarus* **224**, 286-296.

Takai K., et al. (2013) Microbial community development in deep-sea hydrothermal vents in the Earth and the Enceladus. *Am. Geophys. Union Fall Meeting*, Abstract #P53E-09.

Teolis B. D., et al. (2010) Cassini finds an oxygen-carbon dioxide atmosphere at Saturn's icy moon Rhea. *Science* **330**, 1813-1815.

Tobie G., et al. (2008) Solid tidal friction above a liquid water reservoir as the origin of the south pole hotspot on Enceladus. *Icarus* **196**, 642-652.

Tokar R. L., et al. (2012) Detection of exospheric $O_2^+$ at Saturn's moon Dione. *Geophys. Res. Lett.* **39**, L03105, doi:10.1029/2011GL050452.

Tsou P., et al. (2012) LIFE: Life Investigation for Enceladus: A sample return mission concept in search for evidence of life. *Astrobiology* **12**, 730-742.

Turcotte D. L., Schubert G. (2002) *Geodynamics*. Cambridge Univ. Press, New York.

Vance S., Goodman J. (2009) Oceanography of an ice-covered moon. In *Europa* (eds. R. T. Pappalardo, W. B. McKinnon, K. K. Khurana). Univ. of Arizona Press, Tucson, AZ. pp. 459-482.

Vance S., et al. (2007) Hydrothermal systems in small ocean planets. *Astrobiology* **7**, 987-1005.

Venter J. C., et al. (2004) Environmental genome shotgun sequencing of the Sargasso Sea. *Science* **304**, 66-74.

Waite J. H., et al. (2006) Cassini Ion and Neutral Mass Spectrometer: Enceladus plume composition and structure. *Science* **311**, 1419-1422.

Waite J. H., et al. (2009) Liquid water on Enceladus from observations of ammonia and $^{40}Ar$ in the plume. *Nature* **460**, 487-490.

Waite J. H., et al. (2011) Enceladus plume composition. *EPSC-DPS Joint Meeting*, Abstract #61.

Waite J. H., et al. (2013) Enceladus plume composition. *Am. Geophys. Union Fall Meeting*, Abstract #P53E-08.

Westheimer F. H. (1987) Why nature chose phosphates. *Science* **235**, 1173-1178.
58

**Table 1.** Calculated equilibrium speciation of Enceladus' ocean at 0°C and 1 bar, for a nominal concentration of chloride (0.2 molal) and an intermediate concentration of dissolved inorganic carbon (0.07 molal; see Section 2.3). The columns show molal concentrations for three activities of $CO_2$ that span a range consistent with the abundance of $CO_2$ in the plume (see Section 2.2). The constrained activity of $CO_2$ is related to the temperature of the tiger stripes on Enceladus' south polar region via Eq. (2). Several decimal places are shown for numerical completeness, but should not be taken to mean that the chemical composition of the ocean is known to this level of precision.

| Chemical Species | $a_{CO2} = 10^{-11.2}$ ($T_{tiger} = 170$ K) | $a_{CO2} = 10^{-9}$ ($T_{tiger} = 200$ K) | $a_{CO2} = 10^{-7.3}$ ($T_{tiger} = 230$ K) |
|---|---|---|---|
| $Na^+$ | 0.3449 | 0.3194 | 0.3138 |
| $Cl^-$ | 0.1996 | 0.1996 | 0.1996 |
| $CO_3^{-2}$ | 0.04742 | 0.04777 | 0.04512 |
| $NaCO_3^-$ | 0.02253 | 0.02154 | 0.02015 |
| $OH^-$ | 0.02792 | $2.238 \times 10^{-3}$ | $3.078 \times 10^{-4}$ |
| $HCO_3^-$ | $4.003 \times 10^{-5}$ | $5.093 \times 10^{-4}$ | $3.512 \times 10^{-3}$ |
| NaCl | $4.321 \times 10^{-4}$ | $4.060 \times 10^{-4}$ | $4.008 \times 10^{-4}$ |
| NaOH | $2.680 \times 10^{-3}$ | $2.016 \times 10^{-4}$ | $2.736 \times 10^{-5}$ |
| $NaHCO_3$ | $1.501 \times 10^{-5}$ | $1.789 \times 10^{-4}$ | $1.217 \times 10^{-3}$ |
| HCl | $3.291 \times 10^{-22}$ | $4.111 \times 10^{-21}$ | $2.991 \times 10^{-20}$ |
| pH | 13.208 | 12.115 | 11.255 |

**Figure 1.** The concentration of chloride in Enceladus' ocean as a function of the thickness of water ice above the ocean. The curves show results for models with an ocean underlying all latitudes south of the indicated value (e.g., a global ocean corresponds to 90°). The region labeled "Geophysical Constraint" is consistent with Airy compensation depths (e.g., ice thicknesses) inferred by Iess et al. (2014), and with a south polar (LAT = −50°; Iess et al., 2014) or southern hemispheric (LAT = 0°) ocean as suggested by the more active geology of the south compared to the north (Spencer et al., 2009; Spencer & Nimmo, 2013). The region labeled "Geochemical Constraint" indicates the concentration range deduced by Postberg et al. (2009; 2011) for the plume particles. The dashed horizontal line shows the approximate concentration where hydrohalite (NaCl•2H$_2$O) would precipitate from the ocean (from the FREZCHEM model; Marion et al., 2010).

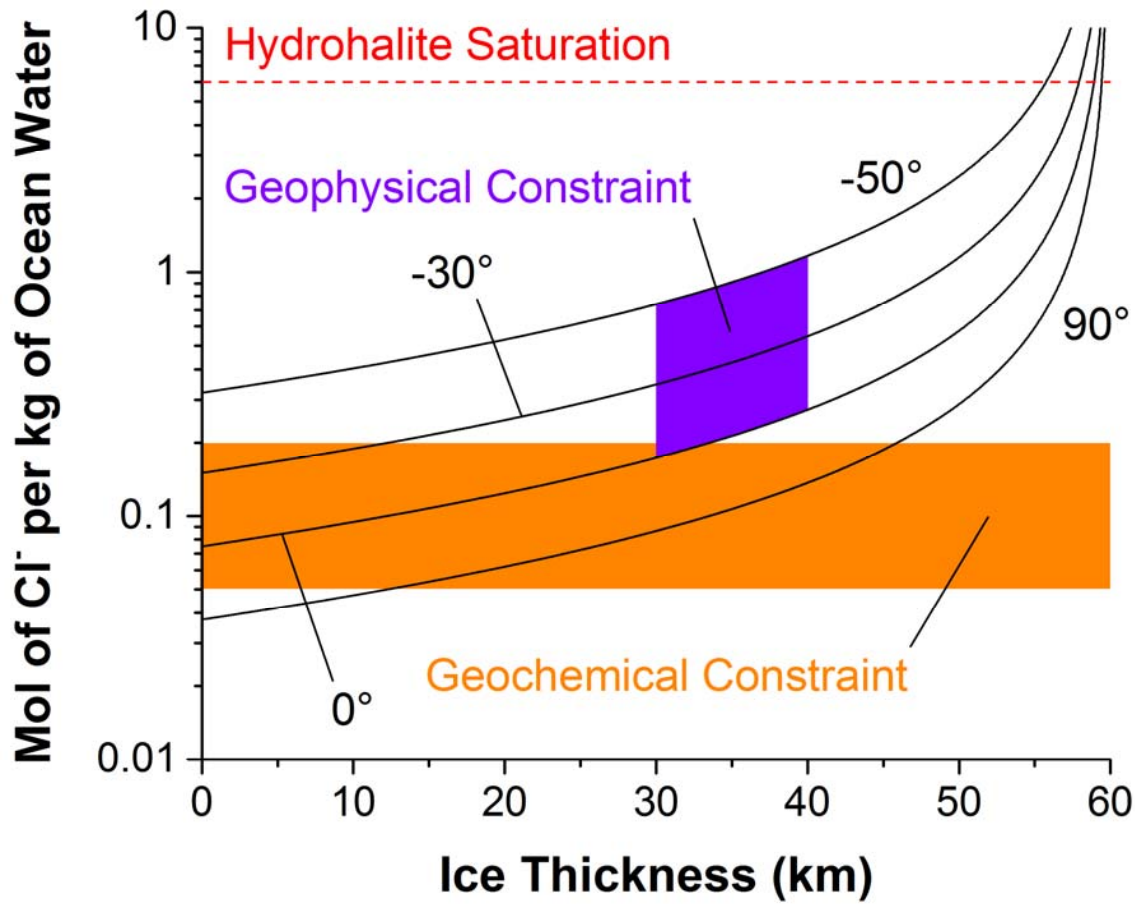

**Figure 2.** The pH of Enceladus' ocean from a simplified model of carbonate speciation at 0°C and 1 bar. The curves show the derived pH at the indicated (molal) concentration of dissolved inorganic carbon. The constraint on the activity of $CO_2$ is from Section 2.2. The orange region shows where the ocean would be for the conservative range in DIC from Section 2.3, while the violet region is consistent with the preferred range. Note that the violet region is overlaying a portion of the orange region. For comparison, the pH of neutral water at this temperature and pressure is 7.47 (Shock et al., 1997).

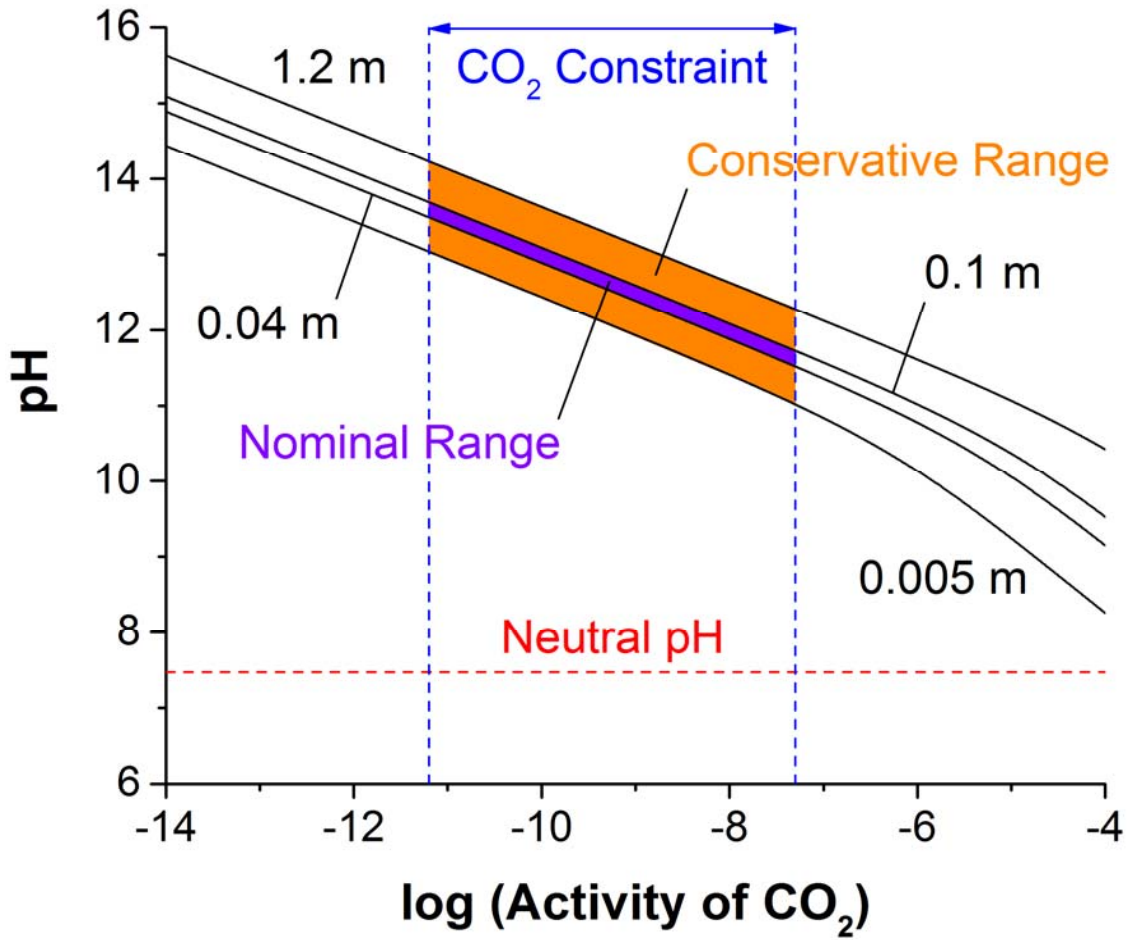

**Figure 3.** The pH of Enceladus' ocean from a detailed model of carbonate speciation at 0°C and 1 bar, and a chloride concentration of (A) 0.05 molal; (B) 0.2 molal; and (C) 1.2 molal. The curves show the derived pH at the indicated concentration of dissolved inorganic carbon. The area between the dashed vertical lines represents the constraint on the activity of $CO_2$ from Section 2.2. The orange region shows where the ocean would be for the conservative range in DIC (DIC/Cl = 0.1-1) from Section 2.3, while the violet region is consistent with the nominal model (DIC/Cl = 0.2-0.5). The dashed horizontal line indicates the pH = 7.47 of neutral water at this temperature and pressure (Shock et al., 1997).

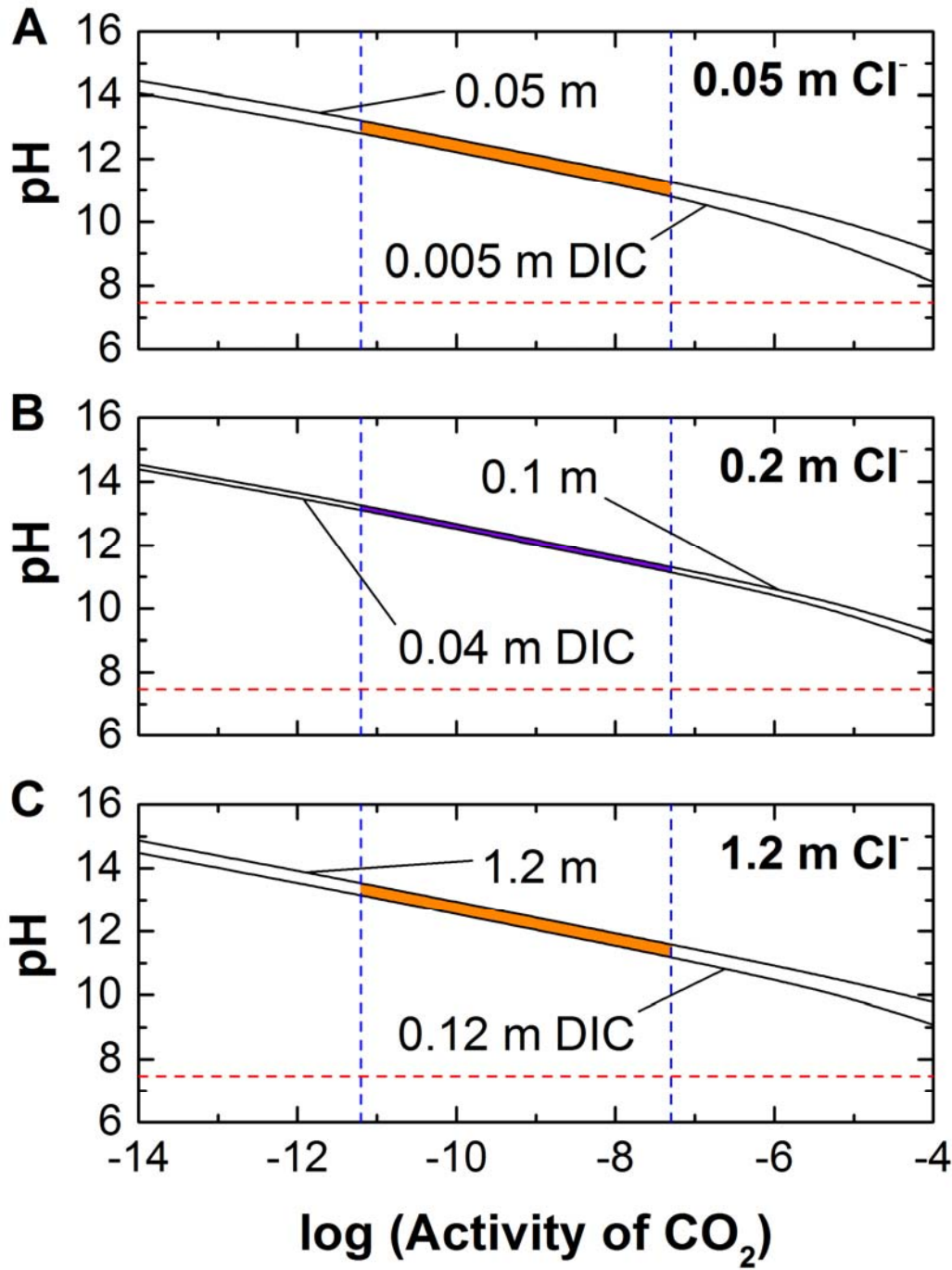

**Figure 4.** Derived concentrations of major chemical species in Enceladus' ocean from a detailed model of carbonate speciation at 0°C and 1 bar, for constraints (see Section 2.3) of (A) 0.05 molal Cl⁻ and DIC/Cl = 0.1; (B) 0.05 molal Cl⁻ and DIC/Cl = 1; (C) 0.2 molal Cl⁻ and DIC/Cl = 0.2; (D) 0.2 molal Cl⁻ and DIC/Cl = 0.5; (E) 1.2 molal Cl⁻ and DIC/Cl = 0.1; and (F) 1.2 molal Cl⁻ and DIC/Cl = 1. Plots (A), (B), (E) and (F) encompass the conservative range in the concentrations of Cl⁻ and DIC, while plots (C) and (D) span the preferred concentration range. The region between the dashed vertical lines represents the constraint on the activity of $CO_2$ from Section 2.2.

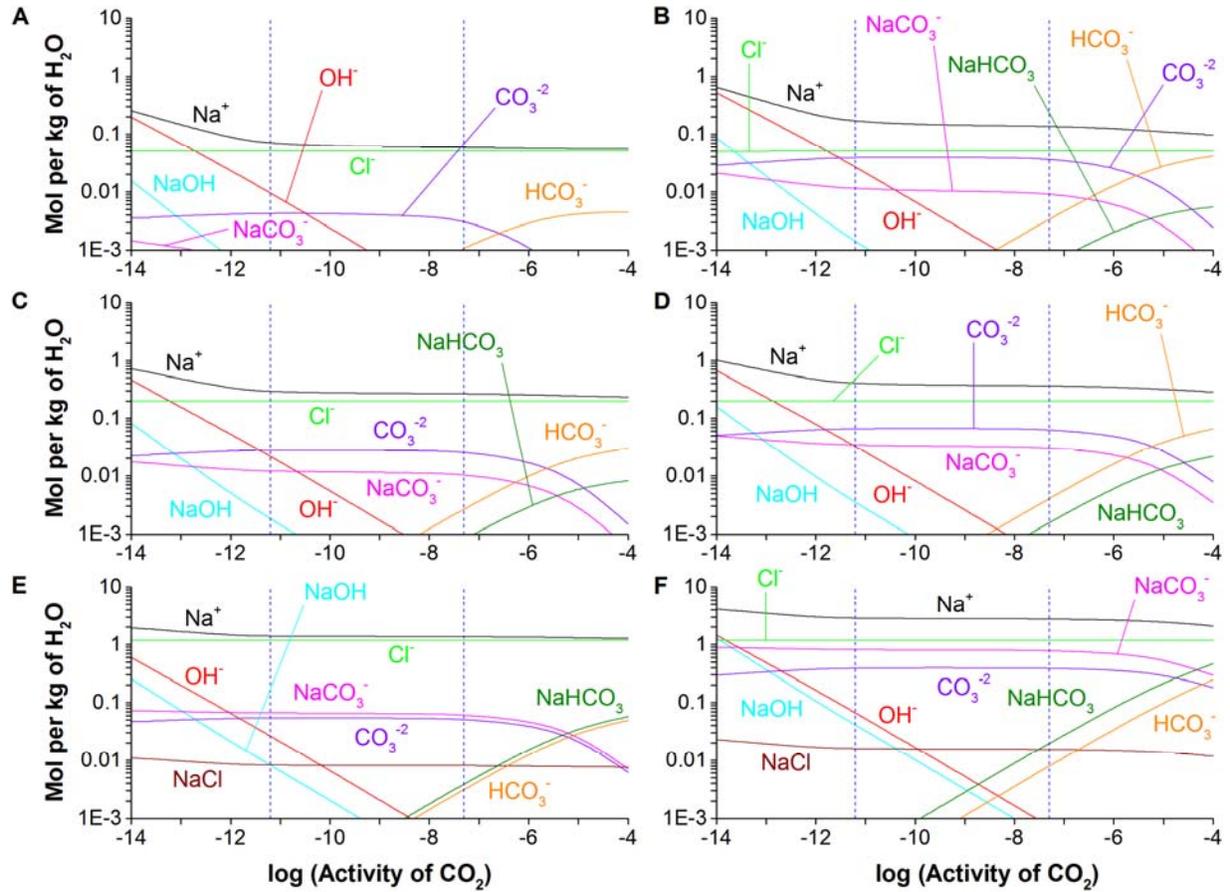